# Existence of Large Room Temperature Ferroelectricity in Chemical Solution Grown PbTiO$_3$ Buffered (BiFeO$_3$)$_{1-x}$-(PbTiO$_3$)$_x$ Films on Pt/Si Substrates

**Ashish Garg[1] and Soumya Kar[1]**

[1]*Department of Materials and Metallurgical Engineering, Indian Institute of Technology Kanpur, Kanpur 208016, India*

**Dinesh Chandra Agrawal[2]**

[2]*Materials Science Programme, Indian Institute of Technology Kanpur, Kanpur 208016, India*

**Shuvrajyoti Bhattacharjee[3] and Dhananjai Pandey[3]**

[3]*School of Materials Science and Technology, Institute of Technology, Banaras Hindu University, Varanasi, India*

Here, we report the existence of large remnant polarization and dielectric constant in PbTiO$_3$ buffered polycrystalline (BiFeO$_3$)$_{1-x}$-(PbTiO$_3$)$_x$ thin films of composition $0.25 \leq x \leq 0.35$ around the morphotropic phase boundary. The films were deposited by chemical solution deposition on platinized silicon substrates. Films reveal a perovskite structure with a tendency to retain rhombohedral structure even at higher PbTiO$_3$ content than that reported in bulk. We observe very large remnant polarization, ~75 μC/cm$^2$, and high dielectric constant, ~2400 at 1 kHz for $x$ = 0.3, which lies in the MPB composition range. The unbuffered films show less than 1 μC/cm$^2$ for the same composition.

Multiferroics exhibiting both ferroelectric and magnetic orderings are very important materials with potential for applications in sensors, actuators and data storage devices.[1,2,3,4] $BiFeO_3$ (BF) is a multiferroic whose room temperature structure belongs to the R3c space group[5,6], and $a^-a^-a^-$ tilt system[7,8], in which the adjacent oxygen octahedra are antiphase rotated about the 3-fold axis of the rhombohedral unit cell. It exhibits high ferroelectric Curie point ($T_C$) of ~ 1100 K[9] and also shows G-type antiferromagnetic behaviour[10] with an incommensurate spiral magnetic ordering[11], below the Neel temperature ($T_N$) of ~643 K.[10] One of the major problems with the films and ceramic samples of BF is its low electrical resistivity, as a result of which the ferroelectric P-E hysteresis loop studies do not reveal the high spontaneous polarization (greater than $50\mu C/cm^2$ reported in epitaxial thin films[12] and single crystals recently.[13,14] Although not quite clear, high leakage of BF is believed to be due to the presence of impurities and due to the reduction of $Fe^{3+}$ to $Fe^{2+}$ because of creation of oxygen vacancies during high temperature sintering.[15,16]

Among various proposed solutions to increase the insulation resistance of BF, formation of solid solution of BF with $ABO_3$ perovskite oxides, such as $PbTiO_3$ (PT)[17] and $BaTiO_3$[18] has been suggested. $(BF)_{1-x}$-$(PT)_x$ (BF-xPT) has been shown to form a continuous solid solution with the existence of a morphotropic phase boundary (MPB) at $x$ = 0.29 (wt%)[19], similar to the well known $Pb(Zr_xTi_{1-x})O_3$ (PZT) ceramics.[20] Further, because of the high Curie point (763K) of PT,[20] compared to other ferroelectric perovskites like $BaTiO_3$ , the Curie point of the mixed system BF-xPT continues to remain high in the range 763 to 1100K. It has recently been shown that BF-xPT system exhibits unusually high tetragonality of ~ 1.187 for tetragonal composition closest to the MPB.[21,22] This is suggested to result from partial covalent character of the Ti/Fe–O and Pb/Bi–O bonds[22] and is an indication of a large achievable remnant polarization. These features of BF-xPT around MPB form a strong motivation for us to carry out studies on the BF-xPT thin films and, in this letter, we report on the observation of large remnant polarization and dielectric constant at room temperature in polycrystalline BF-xPT thin films with compositions near MPB ($x$ = 0.25, 0.30 and 0.35).

BF-xPT films of thickness ~700 nm were synthesized by chemical solution deposition technique. BF-xPT solutions were made by dissolving stoichiometric quantities of high purity bismuth nitrate [$Bi(NO_3)_3.5H_2O$], lead acetate [$(CH_3COO)_2Pb.3H_2O$] and ferric nitrate [$Fe(NO_3).9H_2O$] in acetic acid and 2-methoxyethanol. Titanium butoxide [$TiO(CH_2)_3CH_3)_4$] was stabilized using acetyl acetone and all the solutions were mixed together to yield a clear and stable solution. Prior to the deposition of BF-xPT, platinized Si substrates were coated with a 10 nm thick $PbTiO_3$ (PT) layer to promote the wetting. BF-xPT films were subsequently coated with pyrolysis of each layer at 500°C for 10 min in air followed by post annealing at 750°C for 1 h in air.

Phase analysis of the films was carried out using a Thermoelectron grazing angle X-ray diffractometer with $CuK_\alpha$ radiation ($\lambda$=1.54056 Å). Scanning Electron Microscope (FEI Quanta 200) was utilized to observe the grain morphology. For electrical characterization, Pt electrodes of diameter 0.2 mm were deposited on the films through a shadow mask by sputtering. The dielectric and ferroelectric properties of the films were measured using HP 4192A impedance analyzer and Radiant Precision LC ferroelectric tester while current-voltage characteristics were measured using a Keithley 6517A electrometer. Vibrating sample magnetometer (ADE, EV-7VSM, USA) was used to measure the magnetic characteristics.

Fig. 1 shows the X-ray diffraction patterns of the BF-xPT films of three different compositions ($x$ = 0.25, 0.30 and 0.35) annealed at 750°C. All the peaks in the XRD pattern shown in Fig. 1(a) for the films with composition $x$ = 0.25 could be indexed using a rhombohedral structure. Additional peaks corresponding to the tetragonal phase were observed in the films of composition $x$ = 0.30 and 0.35 (see Fig. 1(b) and (c)), reflecting the onset of the MPB.[22] The observation of pure rhombohedral phase for x = 0.25 and coexistence of rhombohedral and tetragonal phases for x =0.30 is in accordance with the work of Bhattacharjee et al.[22] However the presence of a small amount of rhombohedral phase for $x$ = 0.35 for which the bulk ceramics show pure tetragonal structure[22], suggests that the morphotropic transition in thin films has a wider phase coexistence region than chemically homogeneous bulk ceramics. This could be due to the substrate induced strain in the thin films along with the clamping action of the substrate, posing an impediment to the phase transition, as reported for other perovskite oxide thin films.[23,24] Microstructural analysis of the films using SEM showed that the films were free of any porosity and cracks.

Fig. 2 (a) shows the room temperature ferroelectric hysteresis loops for the BF-xPT films ($x$ = 0.25, 0.30 and 0.35) annealed at 750°C. The measurements were made using a bipolar configuration at 10V and 1 kHz frequency. While the hysteresis loops for $x$ = 0.25 and 0.30 were normal, the loops for $x$ = 0.35 showed large degree of polarization relaxation evident from a large step on the negative side of the polarization axis. Further, the value of Pr is extremely large for the intermediate composition $x$ = 0.30 (75 $\mu C/cm^2$) and drops to lower values for $x$ = 0.25 (~48 $\mu C/cm^2$ ) and $x$ = 0.35 (~57 $\mu C/cm^2$ ). This suggests that the MPB in this films of BF-xPT lies very close to $x$ = 0.30 in excellent agreement with the recent report of the MPB region to lie in the composition range 0.27<x<0.31 in chemically homogeneous BF-xPT bulk ceramics.[22] Polarization values obtained in our films are substantially higher than the remanent polarization values reported in the earlier works, especially those reported at RT.[25,26,27,28,29,30,31,32] Previous studies on chemical solution grown films of BF-xPT have also shown low polarization values (~2 μC/cm$^2$) at room temperature.[26] Significantly higher values (~50 μC/cm$^2$) of remnant polarization in chemical solution grown[27] and pulsed laser deposited films[28,29] of BF-xPT have been reported for measurements carried out at -190°C and -10°C, respectively. We must emphasize here that our measurements were made at a frequency of 1 kHz and at these frequencies, leakage contribution is not as subdued as it is at higher frequencies (5 kHz or higher)[28]: for instance, the 1kHz ferroelectric response of the PLD grown BFPT films at -10°C[29] is much inferior than room temperature measurements of sol-gel derived BFPT films in the present study. It is also known that decreased leakage contribution at low temperatures is responsible for improved ferroelectric behavior, also reported earlier for pure BF thin films.[33]

Although the room temperature ferroelectric behaviour of our samples may not replicate a saturated ideal ferroelectric hysteresis loop, it is far superior than previous reported RT P-E loops of BFPT films. The highest room temperature value of $P_r$ reported previously is ~16 μC/cm$^2$ at 367 kV.cm$^{-1}$ in ZnO buffered 10 at. % lanthanum doped $(BiFeO_3)_{0.6}$–$(PbTiO_3)_{0.4}$ films.[30] In our case, we believe that the dramatic improvement in the room temperature ferroelectric behavior (~70 μC/cm$^2$ at room temperature) is due to the intermediate $PbTiO_3$ layers (~10 nm) between the film and the bottom Pt electrodes. Besides promoting the wetting of the BF-xPT solution to the substrate, PT layers also enhance the ferroelectric properties by providing additional insulation at the interface between the film and the substrate and act as a barrier to the charge flow. The BF-xPT films for x = 0.30 without a PT buffer layer exhibit weak and leaky hysteresis behavior, as reported by earlier workers too[26] (see the inset in top left corner of Fig. 3).

Fig. 2 (b) and (c) show the variation of room temperature dielectric constant and loss with frequency for the BF-xPT films with three different compositions. As the figure suggests, dielectric constant $\varepsilon_r$ at 1 kHz is ~ 3300, ~2400 and ~1930 for $x$ =0.35, 0.30 and 0.25, increasing from 1930 for $x$ = 0.25 to 2400 for $x$ = 0.30. However, it does not decrease for $x$ = 0.35 as expected for MPB systems.[20] This increase does not seem to be due to intrinsic reasons but is most likely due to enhanced conductivity losses, as revealed by higher leakage current density (see Fig. 3) and loss tangent for $x$ = 0.35 (Fig. 2(c)), as compared to those for $x$ = 0.30. (see Fig. 2( c) and Fig. 3). Loss of PbO during annealing of the films at 750°C could be the reason for the enhanced tan $\delta$ and leakage current density. The loss tangent for $x$ = 0.25 and 0.30 is nearly frequency independent for $10^2$ to $10^5$ Hz but starts increasing for higher frequencies. The loss tangents for $x$ = 0.35 is much higher than that for $x$ = 0.25 and 0.30 and starts increasing above $10^4$ Hz. The in decrease in the dielectric constant and increase in the loss beyond ~$10^5$ Hz could also be due to RC/LC resonance during the measurement.

Fig. 3 shows the room temperature leakage characteristic of the three films. Films of composition $x$ = 0.30 showed minimum leakage whilst films with $x$ = 0.25 showed maximum leakage current. The maximum leakage for $x$ = 0.25 can be attributed to higher BF content of the film, whereas the next highest value for $x$ = 0.35 could be attributed to PbO loss during post deposition thermal annealing. The leakage current density of films with composition $x$ = 0.30 is ~0.1 µA/cm$^2$ and increases to 10 mA/cm$^2$ at higher fields, (~ 100 kV/cm). The slope of the log J *vs* log E plots was 2-3 in all the films suggesting non-ohmic mechanism of conduction for the three compositions. Further, log J vs $E^{1/2}$ plot, shown in the inset in Fig.3, exhibits better linearity and is indicative of Schottky or Poole-Frenkel Emission type of conduction mechanism.

Fig. 4 shows the magnetic hysteresis loops for the BF-xPT films of all compositions and one can notice that although the remnant magnetization remains ~1 emu/cc for all films, the films of composition $x$ = 0.30 show maximum value of saturation magnetization, up to ~6 emu/cc. Existence of small but finite remnant magnetization points towards multiferroic nature of the BFPT thin films. However, cannot rule out the possible role of any magnetic impurity phase not detectable by XRD. On the other hand, the magnetic behaviour may as well be intrinsic as saturation magnetization peaks for $x$ = 0.30. Further work is needed to confirm the above.

To summarize, we have shown that chemical solution deposited $PbTiO_3$ buffered polycrystalline $(BiFeO_3)_{1-x}$-$(PbTiO_3)_x$ thin films possess very high values of room temperature remnant polarization (~75 $\mu C/cm^2$) and dielectric constant (~2400) for $x$ = 0.30, a composition corresponding to the morphotropic phase boundary region.[22] The polarization values are the highest achieved so far for BF-xPT films. We believe that the presence of $PbTiO_3$ buffer layer prior to the deposition of BF-xPT thin films helps in improving the properties substantially by promoting the wetting of the BF-xPT solution to the substrate as well as by acting as a barrier for charge conduction. The leakage current densities, however, remain high (~$10^{-2}$ $A/cm^2$ at 10 V). Efforts are in progress to reduce the leakage in these films.

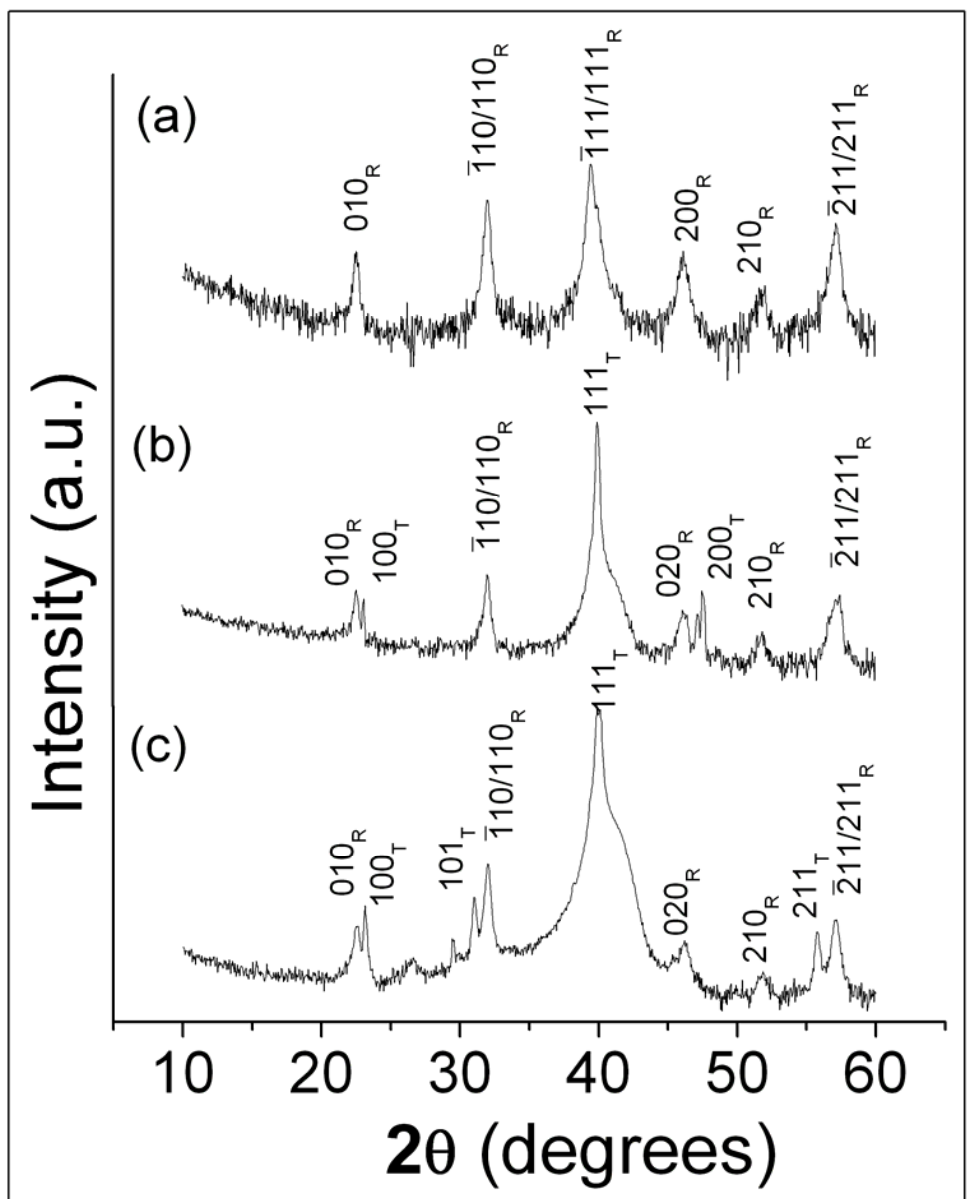

Figure 1      XRD patterns of BF-xPT films with compositions (a) $x = 0.25$, (b) $x = 0.30$ and (c) $x = 0.35$

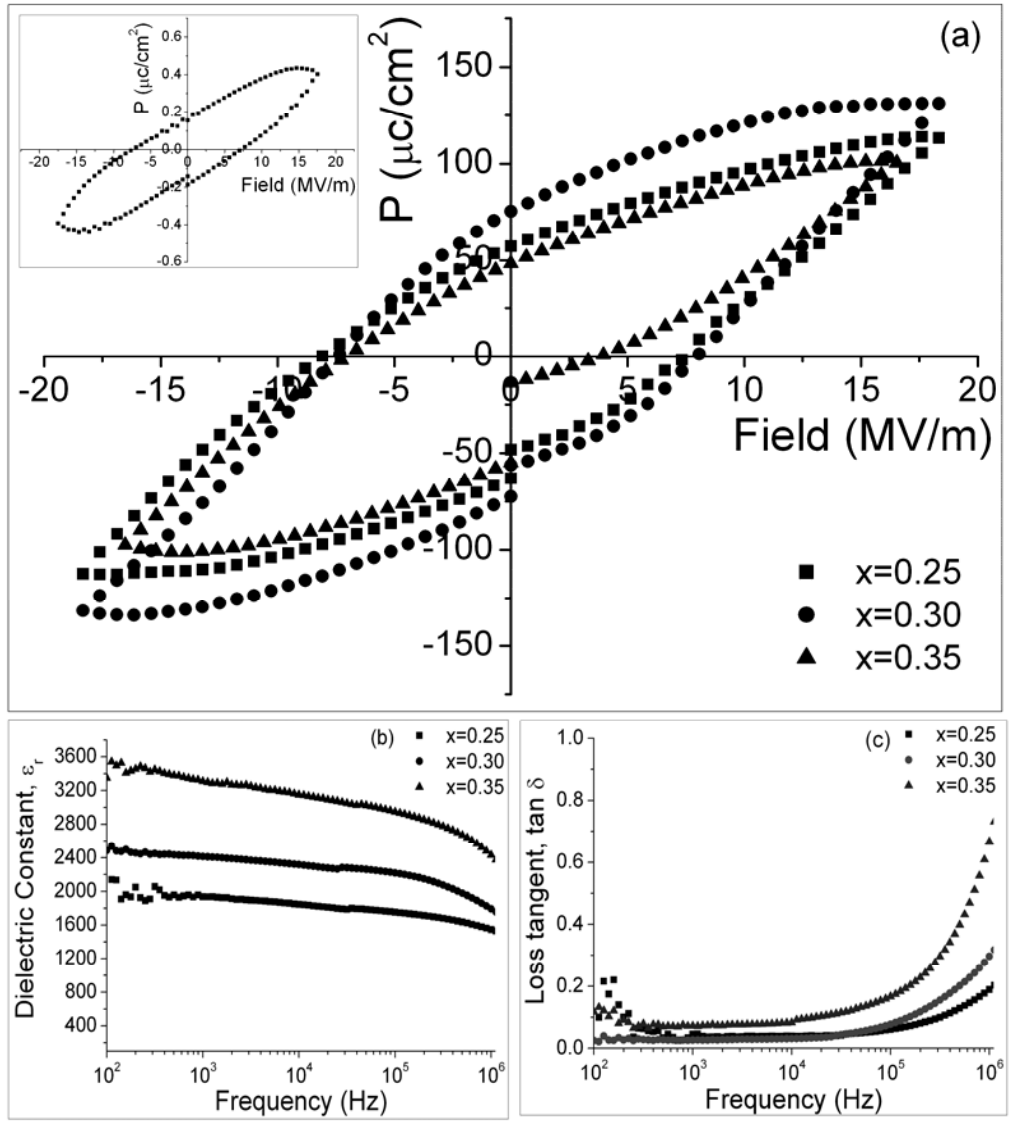

Figure 2     (a) Ferroelectric hysteresis loops at 1 kHz and plots of (b) dielectric constant and (c) loss tangent with frequency for BF-xPT films of compositions $x$ = 0.25, 0.30 and 0.35. Inset in the top left corner of (a) shows the hysteresis behavior of films of $x$ = 0.30 without PT buffer layer. All the measurements were carried out at room temperature.

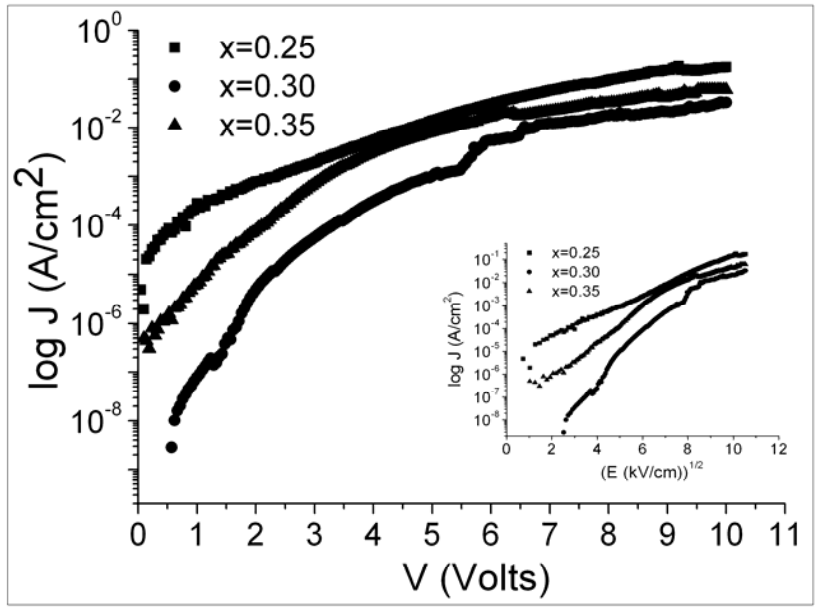

Figure 3    Room temperature leakage characteristics of BF-xPT films of compositions

$x$ = 0.25, 0.30 and 0.35

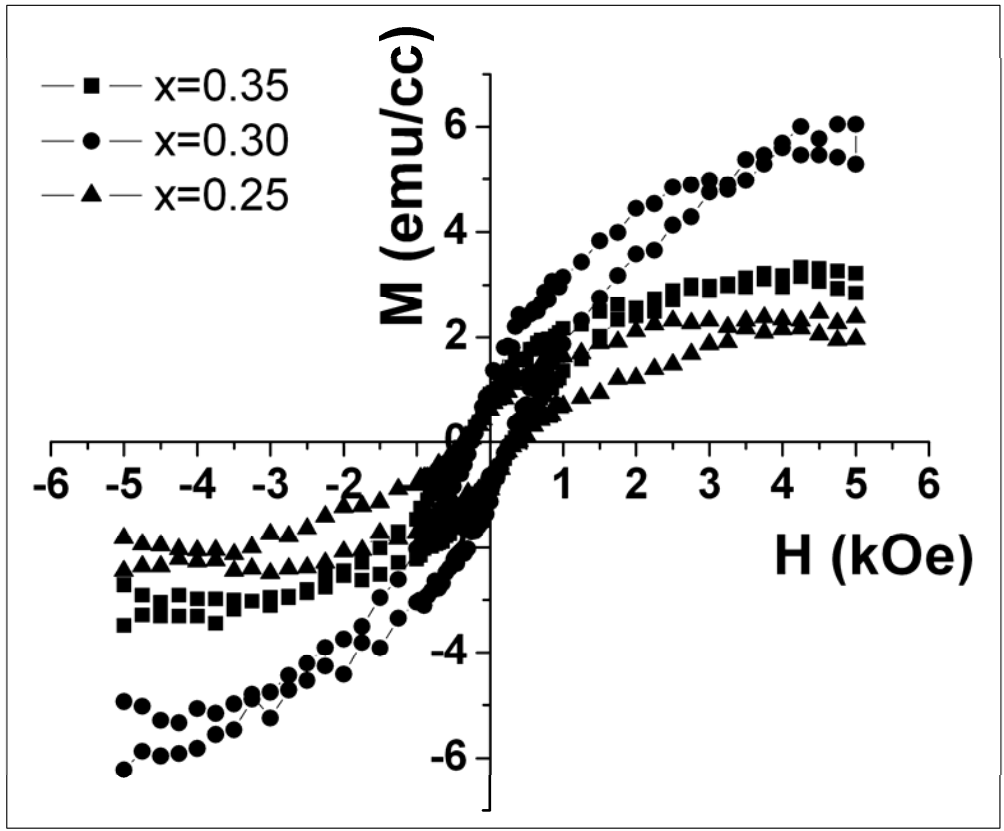

Figure 4 Room temperature magnetic hysteresis (M-H) loops of the BF-xPT thin films of compositions $x$ = 0.25, 0.30 and 0.35; M: magnetization and H: applied field